\documentclass[aps,prl,reprint,superscriptaddress,showpacs]{revtex4-1}

\usepackage{graphicx}
\usepackage[american]{babel}
\usepackage[utf8]{inputenc}
\usepackage[T1]{fontenc}
\usepackage{bbm,ae,aecompl}
\usepackage{amsmath}
\usepackage{amssymb}
\usepackage{amstext}
\usepackage{amsthm}
\usepackage{latexsym}
\usepackage{verbatim}
\usepackage{natbib}
\begin{document}

\title{Orbitally resolved lifetimes in $\text{Ba}(\text{Fe}_{0.92}\text{Co}_{0.08})_2\text{As}_{2}$ measured by ARPES}

\date{\today}
\pacs{79.60.-i, 71.18.-y, 71.30.-h}

\author{V. Brouet}

\author{M. Fuglsang Jensen}
\affiliation{Laboratoire de Physique des Solides, Universit\'{e} Paris-Sud, UMR 8502, B\^at. 510, 91405 Orsay, France}

\author{A. Nicolaou}

\author{A. Taleb-Ibrahimi}

\author{P. Le F\'{e}vre}

\author{F. Bertran}
\affiliation{Synchrotron SOLEIL, L'Orme des Merisiers, Saint-Aubin-BP 48, 91192 Gif sur Yvette, France}

\author{A. Forget}

\author{D. Colson}
\affiliation{Service de Physique de l'Etat Condens\'{e}, Orme des Merisiers, CEA Saclay, CNRS-URA 2464, 91191 Gif sur Yvette Cedex, France}

\begin{abstract}
Despite many ARPES investigations of iron pnictides, the structure of the electron pockets is still poorly understood. By combining ARPES measurements in different experimental configurations, we clearly resolve their elliptic shape. Comparison with band calculation identify a deep electron band with the $d_{xy}$ orbital and a shallow electron band along the perpendicular ellipse axis with the $d_{xz}$/$d_{yz}$ orbitals. We find that, for both electron and hole bands, the lifetimes associated with $d_{xy}$ are longer than for $d_{xz}$/$d_{yz}$. This suggests that the two types of orbitals play different roles in the electronic properties and that their relative weight is a key parameter to determine the ground state. 

\end{abstract}

\maketitle

There is a consensus that the multiband nature of iron pnictides is essential for defining their electronic properties. Many models for the superconducting and antiferromagnetic orders heavily rely on the interaction between different electron and hole Fermi Surface (FS) sheets \cite{MazinPRL08,Paglione:2010p271}. On the experimental side, transport \cite{RullierAlbenquePRL09,FangPRB09}, Raman \cite{ChauvierePRB10,MuschlerPRB09} and quantum oscillations experiments \cite{ColdeaPRL08,ShishidoPRL10} seem to detect predominantly electrons, suggesting they have longer lifetimes than holes. However, ARPES experiments that can image independently hole and electron bands has not evidenced so far a clear difference between them that could explain this behavior. In fact, it has been more difficult to clearly image the electron pockets than the hole pockets with ARPES \cite{ThirupathaiahPRB10,MalaebJPSJ09,LiuNatPhys10}. Recently, an ARPES study resolved them in more details \cite{ZhangPRB11}, but proposed a structure so different from theory, that it clearly calls for more investigation. 

In this paper, we report ARPES measurements in Ba(Fe$_{0.92}$Co$_{0.08}$)$_2$As$_2$, which corresponds to optimal Co doping for superconductivity ($T_c$=23K). Similar features are found for other members of the BaFe$_2$As$_2$ family \cite{JensenMag}. We first clarify that, in appropriate experimental conditions, only the electron pockets of the 1Fe-Brillouin Zone (BZ) \cite{Graser2010} are observed, which greatly simplifies the spectra. The electron pocket is an ellipse made out of two different orbitals with opposite parities and markedly different dispersions. Along one axis, the band is odd and very shallow (50meV), while it is even and much deeper (more than 100meV) along the perpendicular axis. These dispersions correspond very well to band structures renormalized by a factor of 3, suggesting to associate the shallow band with the $d_{xz}$/$d_{yz}$ orbitals and the deep band with the $d_{xy}$ orbital. The parity measured by ARPES for the deep electron band is not that expected for $d_{xy}$, which is due to interferences between the 2 Fe of the unit cell \cite{Brouet1Fe2Fe}. We find that, both for hole and electron pockets, the lifetimes are about twice longer on the $d_{xy}$ parts of the FS than on the $d_{xz}$/$d_{yz}$ parts. These anisotropies are in good agreement with a recent study by Kemper et al. \cite{Kemper2011}. As the $d_{xy}$ parts of the electron pockets are also associated with higher Fermi velocities on the electron pockets, this probably explains why electrons dominate in many experimental techniques. This also suggests that the two different orbitals play different roles in the electronic properties. 

Single crystals were grown using a FeAs self-flux method.  ARPES experiments were carried out at the CASSIOPEE beamline at the SOLEIL synchrotron, with a Scienta R4000 analyser, an angular resolution of $0.2^{\circ}$ and an energy resolution better than $10$ meV. Band structure calculations were perfomed within the local density approximation, using the Wien2K package \cite{Wien2k} and the experimental structure of BaFe$_2$As$_2$.

Before presenting our ARPES data, we recall two particularities of the BZ structure that will be important for our discussion \cite{Graser2010,Andersen2011}. First, as in all iron pnictides, the BZ (thick red lines in Fig. 1) corresponds to a unit cell with 2 Fe, because of the inequivalent positions of As above and below the Fe plane. Nevertheless, it is sometimes useful to think in an “unfolded” BZ containing just 1 Fe (dotted blue line), because this divides the number of bands by two. Moreover, folded bands typically have a very small intensity in photoemission (it is proportional to the strength of the potential at the origin of the folding) and they may not be detected \cite{VoitScience2000,brouet:2008,Brouet1Fe2Fe}. Second, and this is specific to the stacking of the FeAs plane in BaFe$_2$As$_2$ \cite{Paglione:2010p271}, the folding takes place between different $k_{z}$, namely along the vector q=(1,1,1) shown in the inset of Fig. 1. Following ref. \cite{Graser2010}, we sketch with solid lines the electron pockets expected in the unfolded BZ and with dotted lines the folded ones. The electron pockets are ellipses made out of of $d_{xy}$ and $d_{xz}$/$d_{yz}$ orbitals with major axis oriented towards Z, which rotates as a function of $k_{z}$. This structure is completely at variance with that proposed in ref. \cite{ZhangPRB11}, where $d_{x^2-y^2}$ would form a large square pocket and $d_{xz}$ and $d_{yz}$ an ellipse.

\begin{figure}[tbp]
\centering
\includegraphics[width=0.45\textwidth]{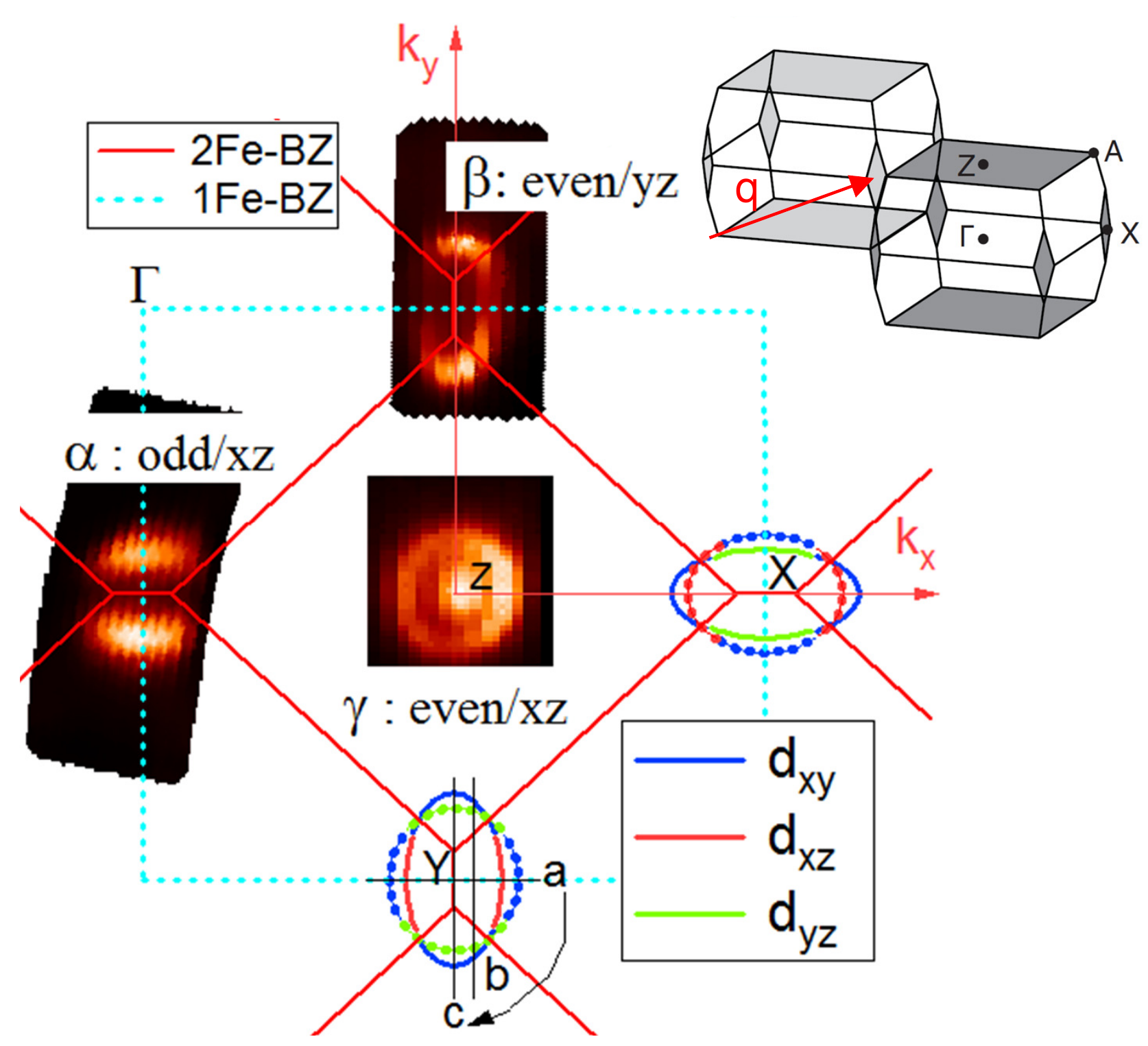}
\caption{ FS measured at T=25 K, $\hbar\omega$=34eV and polarization along $k_y$, except for the hole bands at Z, where it is in the $xz$ plane. The plane of light incidence is $xz$ and the analyser slits are fixed along $k_y$. Ellipses at BZ corners are sketches of the electron pockets expected in theory (adapted from ref. \cite{Graser2010}). Main bands are shown with solid lines and bands folded in the 2Fe-BZ by dotted lines. Orbital characters are coded by colors. Top inset shows the stacking of the 3D BZ.}
\label{PM-figure}
\end{figure}

Fig. 1 presents the FS at 25K. The photon energy $\hbar\omega$ is chosen so that normal emission (Z) corresponds to $k_z$=1 modulo $\pi$/c, where c=6.5\AA~is the distance between 2 FeAs planes \cite{BrouetPRB09,supplementary}. Hole pockets are found at the 2Fe-BZ center and electron pockets at its corners. The polarization defines the symmetry of the orbitals that can be observed. To select even or odd orbitals with respect to a plane of symmetry, the polarization must be even or odd with respect to that plane \cite{MansartPRB11,ZhangPRB11}. Usually, electrons are detected in the plane of light incidence (here, $xz$) and polarization is chosen either in this plane to detect even orbitals (so called p-polarization, as for the hole pockets in panel $\gamma$) or perpendicularly to it to detect odd orbitals (so called s-polarization, as for the electron pockets in panel $\alpha$). However, the intensity for electron pockets in p-polarization turns out to be extremely low here. Instead, we have found that keeping polarization along $k_y$ and detecting electron in the $yz$ plane (panel $\beta$), which is an alternative way to detect even orbitals \cite{Asensio:2003}, yields a much larger intensity. Indeed, it allows to detect in panel $\beta$ parts of the electron pockets oriented towards the zone center that are missing in most measurements reported so far \cite{ThirupathaiahPRB10,MalaebJPSJ09,LiuNatPhys10}, except ref. \cite{ZhangPRB11}.

\begin{figure}[tbp]
\centering
\includegraphics[width=0.5\textwidth]{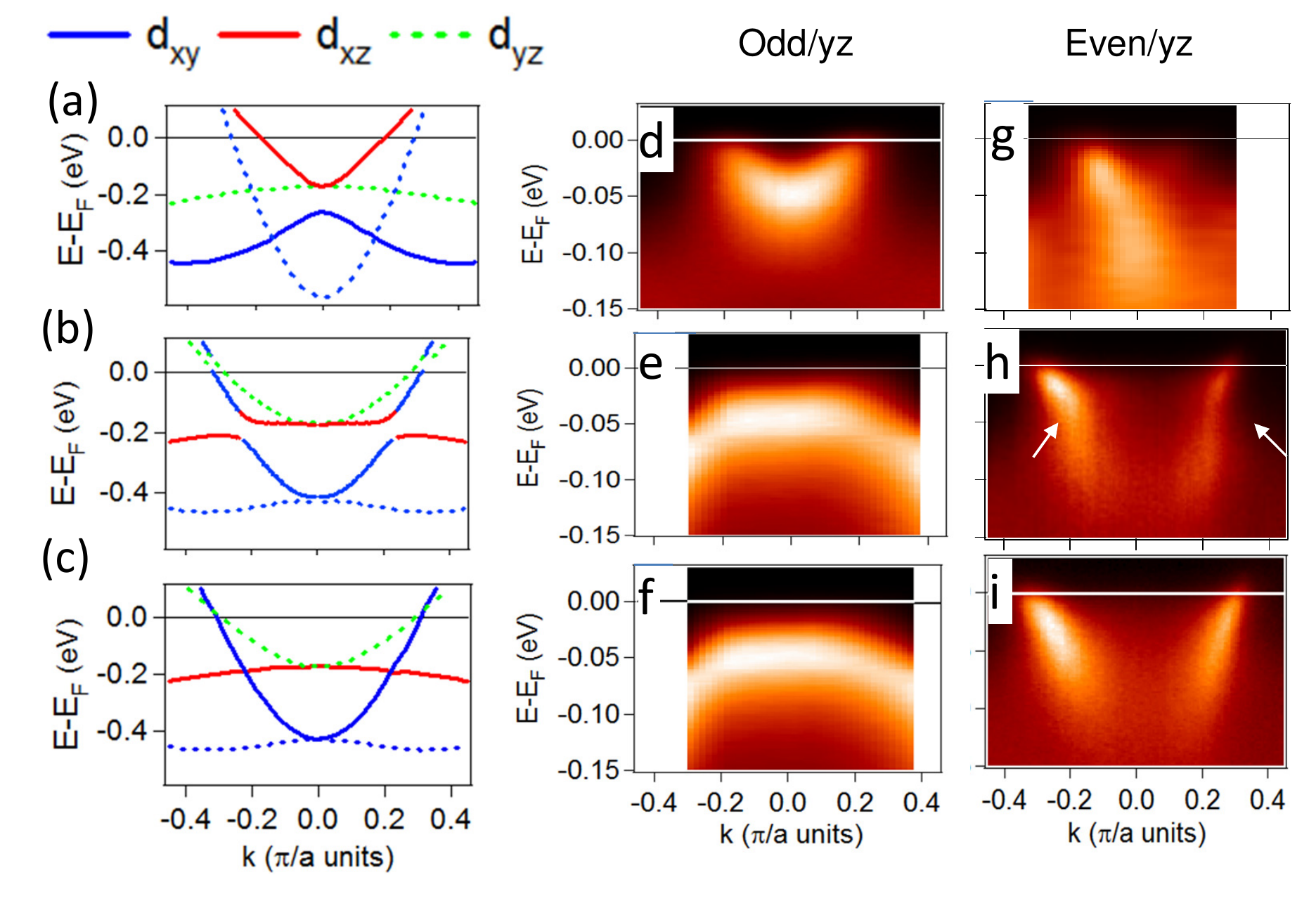}
\caption{(a-c) Band structure calculated for the 3 direction indicated in Fig. 1 as a (ellipse minor axis), b (parallel to major axis, but distant by 0.05$\pi$/a) and c (major axis). Color indicate the orbital character and dotted lines the folded bands in the 2Fe-BZ. (d-f) Corresponding ARPES intensity plots for odd geometry (panel $\alpha$ rotated by 90$^\circ$). (g-i) Same for even geometry (panel $\beta$).}
\label{PM-figure}
\end{figure}

 We now examine in more details how many bands form these FS and how they connect together. Fig. 2 compares the measured and calculated dispersions in the 3 directions (a,b,c) indicated at bottom of Fig. 1. As odd measurement, we use those of panel $\alpha$ rotated by 90$^\circ$, which are equivalent by symmetry. We indicate by colors their main orbital character, although bands are frequently hybridized here in a more complex way \cite{Andersen2011}. For pure atomic orbitals, $d_{xy}$ and $d_{xz}$ would be odd with respect to the $yz$ plane and $d_{yz}$ even. Note that $d_{xz}$ and $d_{yz}$ switch roles along $k_x$ and $k_y$, but their parity with respect to $xz$ and $yz$ also switches, so that the selection rules remain identical. In the odd geometry (d-f), only one band is clearly observed. It corresponds to $d_{xz}$ in the calculation (red line), which complies with the selection rule. It forms a shallow electron band of 50meV along the minor axis of the ellipse (Fig. 2d) then becomes more squarish (Fig. 2e) and finally transforms into a band of opposite curvature 50meV below $E_F$ (Fig. 2f). At $k_{z}$=0, this band is expected to stretch to form the ellipse long axis (see sketches in Fig. 3). In Fig. 3(b) and supplementary information \cite{supplementary}, we show that this is actually well observed experimentally. 

The $d_{yz}$ band (green line) is not seen in this odd measurement, as expected from symmetry. On the other hand, the $d_{xy}$ band should be observed, but is not. Surprisingly, it is the band detected in Fig. 2(i) in even geometry that perfectly matches the dispersion of $d_{xy}$ [see also Fig. 4(b)]. It is a much deeper band with higher Fermi velocities. It clearly does not fit with $d_{yz}$, which should cross $E_F$ at similar points, but is much shallower, as clearly observed for its symmetric in odd geometry, both at $k_{z}$=0 and $k_{z}$=1 \cite{supplementary}. To convince oneself that this band is really $d_{xy}$, it is interesting to follow how it transforms into the shallow band. Fig. 2(b) shows that as soon as one goes away from the diagonal, a gap opens between the $d_{xy}$ and $d_{xz}$ bands. The upper part of the deep band progressively transforms into the shallow band near $E_F$, while the $d_{xy}$ character is concentrated at the bottom of the band. This behavior is clearly observed in the data of Fig. 2(h), with a “kink” (white arrow) appearing in the dispersion of the deep band, where it crosses $d_{xz}$ that is not observed in this geometry. In Fig. 2(g), the intensity is quite weak and unclear, but probably contains residual contribution from $d_{xz}$, which is not strictly forbidden outside the $yz$ plane \cite{supplementary}. As the dispersion is intimately linked to the shape of one orbital, we conclude that the deep electron band is $d_{xy}$. The problem of parity is solved when one takes into account the 2 Fe of the unit cell \cite{Brouet1Fe2Fe}. Fig. 3 shows that it also follows qualitatively the expected behavior as a function of $k_z$ (its diameter shrinks away from $k_z$=7$\pi$/c=1[$\pi$/c]), although it strongly loses intensity near $k_z$=6$\pi$/c=0[$\pi$/c].

\begin{figure}[tbp]
\centering
\includegraphics[width=0.5\textwidth]{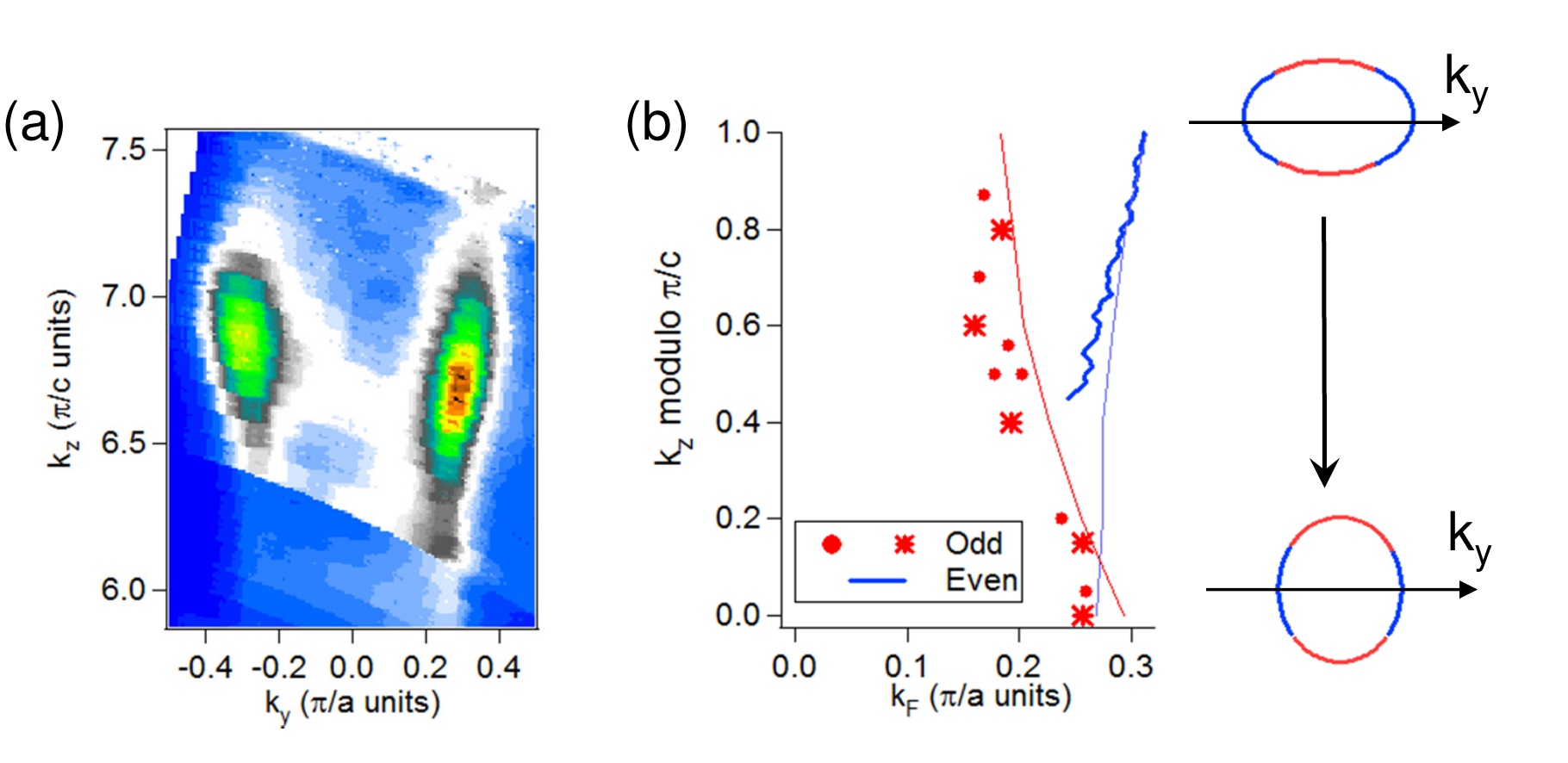}
\caption{(a) intensity integrated at E$_F$ for the deep electron band, for 25<$\hbar\omega$<45eV (see \cite{supplementary} for translation into $k_{z}$). (b) $k_F$ for the deep (blue, extracted from (a)) and shallow (red) electron bands at different k$_z$. For the shallow electron band, two samples were measured for 20<$\hbar\omega$<45eV (circles) and 40<$\hbar\omega$<90eV (stars). Thin lines are guides for the eyes. Ellipses on the right sketch how electron pockets evolve with $k_{z}$.}
\label{PM-figure}
\end{figure}

The structure of the electron pocket is then quite simple, with only two interacting bands, whose dispersions are in remarkable agreement with theoretical predictions. A significant advantage of working with the electron bands is that the dispersion of $d_{xy}$ and $d_{xz}$/$d_{yz}$ are very different, so that they are easy to distinguish on this basis alone, without relying on the polarization dependence that may be complex to analyse in these systems and finally sometimes misleading \cite{ZhangPRB11,supplementary}. This is not the case for the hole bands, where all dispersions are similar [see Fig. 4(a)] and their ordering is very sensitive to details of the calculation, making their assignment very ambiguous. There is a general consensus in the ARPES community that the inner hole band is doubly degenerate and of $d_{xz}$/$d_{yz}$ symmetry, while the origin of the outer hole band is more debated \cite{MansartPRB11,ZhangPRB11,ThirupathaiahPRB10}. It should be $d_{xy}$ in calculations, but mainly appears as even in ARPES and is stronger near $k_{z}$=1. This behavior now resembles that of the deep electron band, suggesting they have the same orbital origin. In the following, we refer for simplicity to these two bands as $d_{xy}$, although they certainly have significant contributions from other orbitals (already roughly 20\% in calculations). 

\begin{figure}[tbp]
\centering
\includegraphics[width=0.5\textwidth]{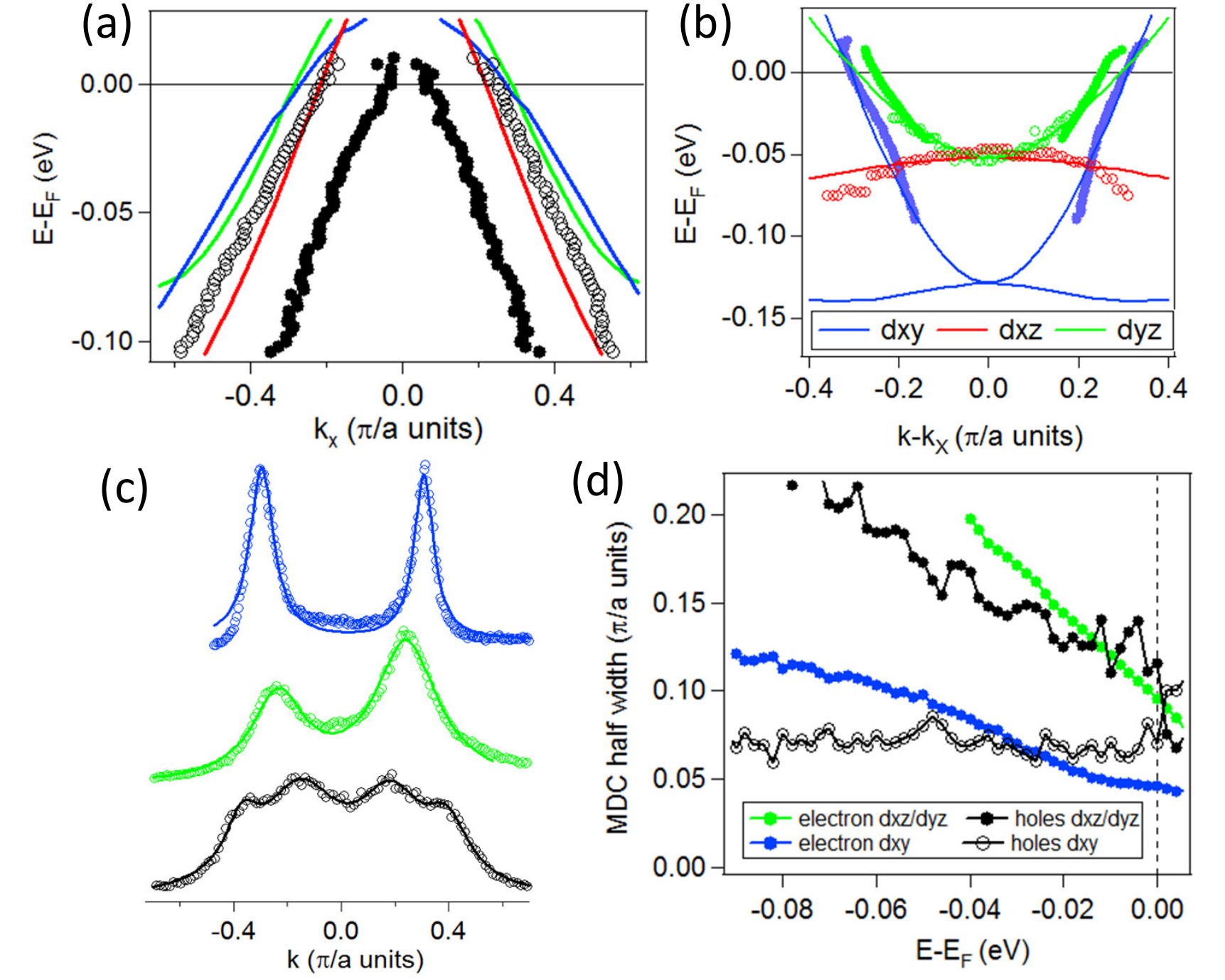}
\caption{(a-b) Dispersion of the holes and electrons bands at $k_{z}$=1, compared with LDA band calculations at $k_z$=0.5 for BaFe$_2$As$_2$, renormalized by a factor 3.3. The open circles in (b) show maximum of EDCs. (c) Representative examples of MDC fits at -0.03eV for holes (bottom), at E$_F$ for shallow (middle) and deep (top) electron bands. (d) Half width of the fitted lorentzians as function of the binding energy, averaged on the left and right bands.}
\label{PM-figure}
\end{figure}
\begin{table}[b]
\caption{\label{''Table 1''} Properties of the different bands at $k_{z}$=1, extracted by the MDC fits of Fig. 4. For the $d_{xy}$ electron band, values at low ($v_{low}$) and high ($v_{high}$) binding energies are quoted.}
\begin{tabular}{|c|c|c|c|c|}  
\hline
      & $k_F$ ($\pi$/a) & $v_F$ (eV.\AA) & m*/m$_b$ &$\delta$k($E_F)$ ($\pi$/a) \\
  \hline
  hole $d_{xz}$/$d_{yz}$ &0.06&0.5&2.7&0.11\\
hole $d_{xy}$ & 0.22&0.4&2.3&0.07\\
electron $d_{xz}$/$d_{yz}$ &0.25&0.6&2.4&0.09\\
 electron $d_{xy}$ &0.30&0.7/1.2&5/2.9&0.04\\
   \hline
\end{tabular}
\end{table}

This clarified, we now show in Fig. 4 the dispersions and widths of the different bands at $k_{z}$=1 (for the shallow band, we use the equivalent of Fig. 2(a) at $k_z$=0 \cite{supplementary}). Some representative examples of the fits used to extract these quantities are shown in Fig. 4(c). Fermi velocities and linewidth are quoted in Table 1. Note that the deep electron band is quite narrow near $E_F$ ($\Delta$k=0.04$\pi$/a, i.e. an inverse lifetime $\hbar$/$\tau$=20meV after multiplication by v$_F$). This is just 1.5 larger than the outer hole band in LiFeAs that was claimed to be a particularly favorable case \cite{KordyukLiFeAs}. This invalidates the idea that lines would be extremely broadened by Co doping in BaFe$_2$As$_2$ \cite{WadatiPRL10} and too broad to perform meaningful self-energy analysis.

In Fig. 4(a) and 4(b), the bands are compared with our LDA calculation at $k_z$=0.5 renormalized by a factor 3.3. This order of magnitude is similar to those measured \cite{YiPRB09,ShishidoPRL10} or calculated \cite{AichhornPRBFeSe} for related systems. The bands were not shifted, although the calculation was done for BaFe$_2$As$_2$. We do observe shifts of the bands between BaFe$_2$As$_2$ and this compound, consistent with electron doping \cite{BrouetPRB09}, but some orbital and $k_{z}$ dependent shifts would be needed to adjust the band structure to experimental data and this is not the scope of this paper. Calculations within DMFT give effective orbital masses of 2.4 for $d_{xy}$ and 2.25 for $d_{xz}$/$d_{yz}$, slightly different from our band masses, but the overall agreement of the calculated correlated band
structure with experimental data is very good. 

Fig. 4(d) shows that the deep electron band is a factor 2 narrower near $E_F$ than the shallow one and that its linewidth increases much less steeply with binding energy. Interestingly, a similar anisotropy is observed for the hole bands having the same orbital characters. Although the hole bands are more difficult to separate [see Fig. 4(c)], we have found the same dependence for the inner band, when the outer band is suppressed by polarization, confirming this behavior. This suggests that the two orbitals play a different role in the electronic structure. This difference is not reproduced in DMFT calculations \cite{AichhornDMFT}, which predict similar scattering rates for these two bands. Considering the importance of spin fluctuations in these systems, it is tempting to attribute the larger linewidth observed for $d_{xz}$/$d_{yz}$ band to a stronger coupling of this band with spin fluctuations than for the $d_{xy}$ band. One could imagine a situation where local moments would form preferentially in this band, while $d_{xy}$ would be principally responsible for the metallicity. Recently, Kemper et al. predicted a strong anisotropy of the lifetimes between $d_{xy}$ and $d_{xz}$/$d_{yz}$ \cite{Kemper2011}, which appears qualitatively very similar to this finding. However, this anisotropy depends sensitively on the nesting between the different bands and, as their model did not include a $d_{xy}$ hole band, it cannot be directly compared.  

Below 40meV, the dispersion of $d_{xy}$ is more strongly renormalized and come closer to that of the $d_{xz}$/$d_{yz}$ band (see Table 1). Simultaneously, the linewidth nearly saturates. This behavior is of course reminiscent of a "kink", which is often attributed to a coupling with a bosonic mode, phonons or magnons, and was sometimes proposed for iron pnictides \cite{KordyukLiFeAs}. Note, however, that the coupling would be very large here $\lambda$=v$_{high}$/v$_{low}$-1=0.7, making this interpretation somewhat unlikely. In this multiband system, we rather believe that it is the signature of interactions between the different bands. Along the diagonal, a direct interaction, as in Fig. 2(h), is not expected. However, 40meV corresponds very well to the $d_{xz}$/$d_{yz}$ bandwidth, suggesting that scattering between electrons in the two bands may take place below this energy scale and tend to homogenize their properties. 

To conclude, we have detailed how $d_{xy}$ and $d_{xz}$/$d_{yz}$ combine to form electron pockets in Ba(Fe$_{0.92}$Co$_{0.08}$)$_2$As$_2$, reconciling ARPES and theory. We have then studied the self-energy for the different orbitals. It reveals a clear anisotropy between the behavior of the $d_{xy}$ and $d_{xz}$/$d_{yz}$, both for the hole and electron bands, with longer lifetimes on $d_{xy}$. As $d_{xy}$ form the extremities of the electron pocket, where v$_F$ is also higher, these parts are likely a "cold spot" on the FS that could dominate some electronic properties, such as transport. We tentatively attribute the shorter lifetimes of $d_{xz}$/$d_{yz}$ to stronger spin fluctuations. We also evidence inter-scattering effects between $d_{xy}$ and $d_{xz}$/$d_{yz}$ near the Fermi level. As they are probably very dependent on the relative nesting of these two bands, this could be a key to stabilize magnetism or superconductivity. To test this idea, following precisely the fate of $d_{xy}$ under isovalent and aliovalent substitutions should be very instructive. 

We thank M. Aichhorn, S. Biermann, S. Graser and P. Hirshfeld for useful discussions. Financial support from the French RTRA Triangle de la physique and the ANR \lq\lq{}Pnictides\rq\rq{} is acknowledged. 

\bibliography{Bib}

\begin{widetext}
\vspace{10mm}
{\large\bf Supplementary information}

\vspace{10mm}

{\bf Fermi Surface at $k_z$=0}
\vspace{3mm}

In Fig. \ref{FS_kz0}, we show the FS in similar conditions as those of Fig. 1, but at a photon energy of 27eV, corresponding approximately to $k_z$=0 on the electron pockets. The real trajectories in a $yz$ plane at 27 and 34eV are sketched in Fig. \ref{FS_kz0}b, using $k_{z}=\sqrt{0.512^2(\hbar \omega-W+V_0)-k_{//}^2}$, where $k_{//}$ is component of the momentum k in the surface plane, W is the work function (4.2eV) and $V_0$ an inner potential estimated to be 14eV [26].

\begin{figure*}[h]
\centering
\includegraphics[width=0.7\textwidth]{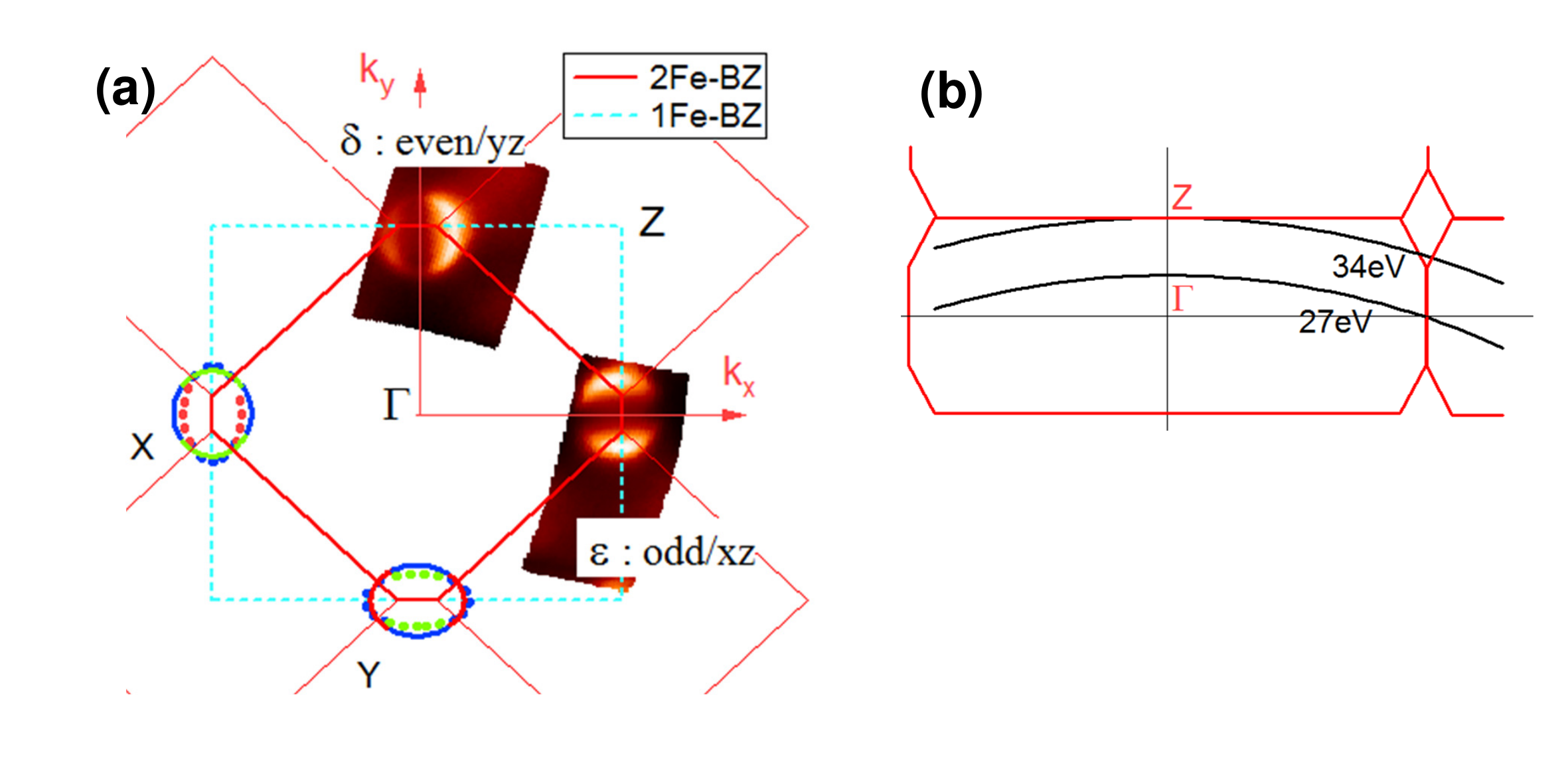}
\caption{(a) FS in same condition as Fig. 1, except the photon energy was 27eV. (b) Sketch of the BZ in the $yz$ plane. Black lines indicate trajectories in k-space for the two photon energies used in this study.}
\label{FS_kz0}
\end{figure*}

In odd geometry, the FS at $k_z$=0 and $k_z$=1 have similar appearances, except that the FS at $k_z$=0 is stretched in the $k_y$ direction. As detailed in Fig. \ref{Odd}, they are both created by the shallow electron band, which stretches, almost as expected theoretically (see also Fig. 3 and ref. [12]). The data in Fig. 4 for the shallow electron band correspond to Fig. \ref{Odd}d. The diameter of the band is somewhat smaller than in theory, for all $k_z$. As one can expect larger electron pockets for this Co doped compound than in pure BaFe$_2$As$_2$, for which the calculation was performed, this is a significant difference. However, the tendency of having smaller pockets than predicted theoretically was already noted before (see for example ref. 20) and probably corresponds to an upward shift of these bands. At 39eV, one starts to distinguish the two electron bands (see below). 

\begin{figure*}[h]
\centering
\includegraphics[width=0.75\textwidth]{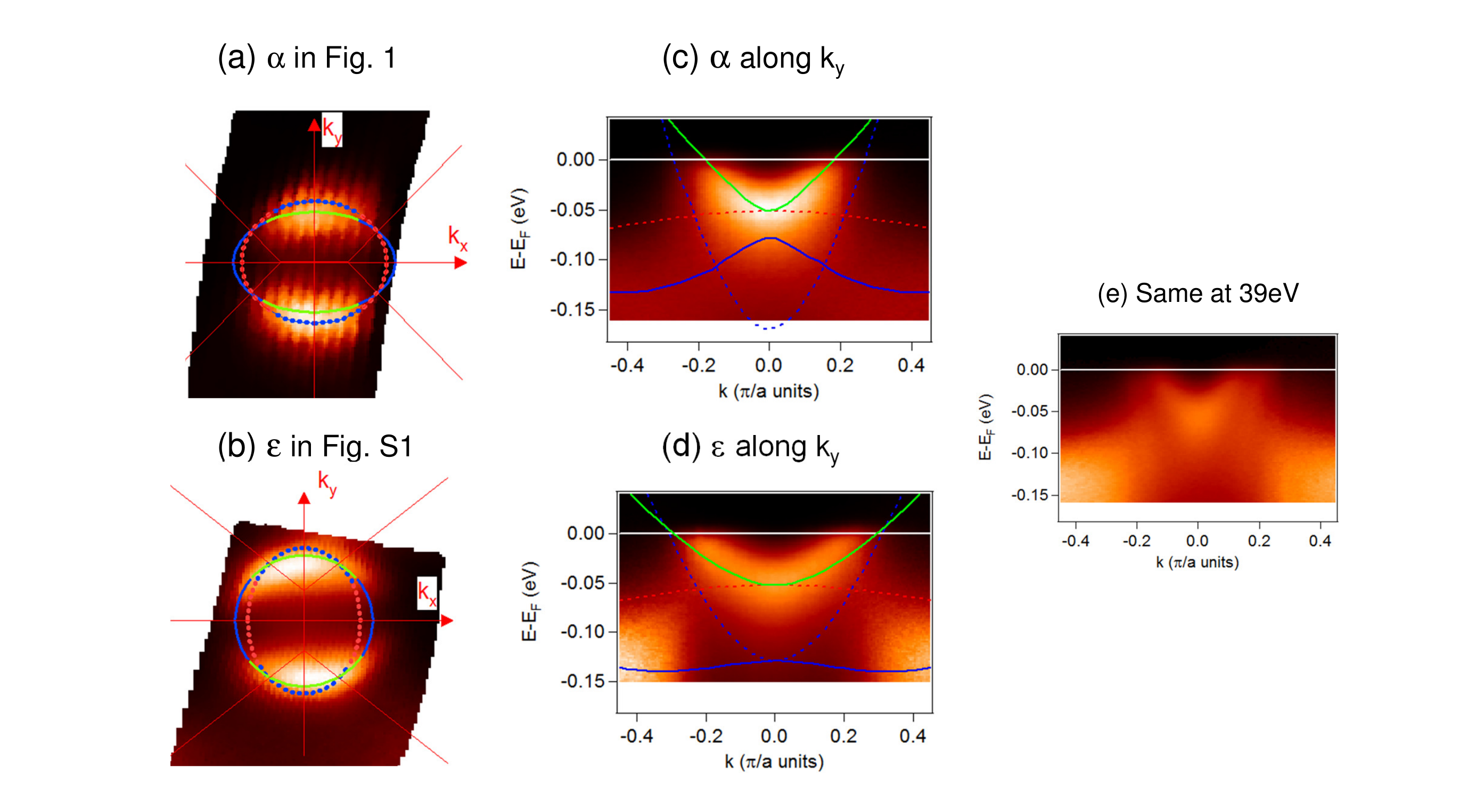}
\caption{(a-b) Zoom on the electron pockets odd/$xz$ of panels $\alpha$ and $\epsilon$ (Fig. 1 and \ref{FS_kz0}), together with sketches of the electron pockets. (c-d) Corresponding dispersions along $k_y$, together with calculated bands rescaled by a factor 3. Blue lines correspond to $d_{xy}$, red lines to $d_{xz}$ and green lines to $d_{yz}$. Solid and dashed lines correspond to main and folded bands. (e) Same at 39eV.}
\label{Odd}
\end{figure*}

In even geometry, the evolution is more complicated, as one would also expect the FS to keep a similar appearance (in this case, it should shrink near $k_z$=0), but the high intensity parts seem to rotate. We have observed a similar FS at the next $k_z$=0 point (48eV), ensuring that it is a periodic effect. A similar behavior was also observed in ref. [12] and it was proposed that the high intensity parts in panels $\beta$ and $\delta$ form a large square pocket of $d_{x^2-y^2}$ symmetry. 

In Fig. \ref{Even}, we present the dispersions along $k_x$ and $k_y$ to better understand the evolution. In Fig. \ref{Even}c, we observe the deep electron band, corresponding well to $d_{xy}$ in the calculation. As shown in Fig. 3a, it loses intensity at lower $k_z$ and it is indeed not detected anymore at $k_z$=0 (Fig. \ref{Even}d). Another band is observed that does not cross $E_F$. It could correspond to $d_{yz}$, but this is unlikely as it is not allowed by symmetry. It is rather the folded $d_{xy}$ that would be shifted up by about 50meV compared to the calculation. Hence, this is as if intensity had switched from the main bands at $k_z$=1 to the folded bands at $k_z$=0. At least, there are strong modulations of the intensity with $k_z$ that are not expected by orbital symmetry alone. 
	
In Fig. \ref{Even}e, the overall intensity is much weaker and more difficult to interpret. We probably observe the main $d_{xy}$ band at high binding energies and some traces of the $d_{yz}$ band near $E_F$. Although, this band is not allowed by symmetry, this is strictly true only in the $yz$ plane (i.e. for the cut along $k_y$) and the intensity is quite weak anyway, so that even a small misalignment could produce this residual intensity. Having observed this, one realizes that the dispersion in Fig. \ref{Even}f is in fact very similar to that in \ref{Even}e, maybe with some more intensity of the folded $d_{xy}$ band. Therefore, the different appearance of the FS is not at all due to a rotation, but to a different contrast in the intensity of the bands along $k_x$ and $k_y$, resulting mainly from the loss of intensity of the deep electron near $k_z$=0. For this reason, we believe these two parts are unlikely to form a large square pocket of $d_{x^2-y^2}$ symmetry, as proposed in ref. 12, because they are not the same bands. 
\begin{figure*}[h]
\centering
\includegraphics[width=0.75\textwidth]{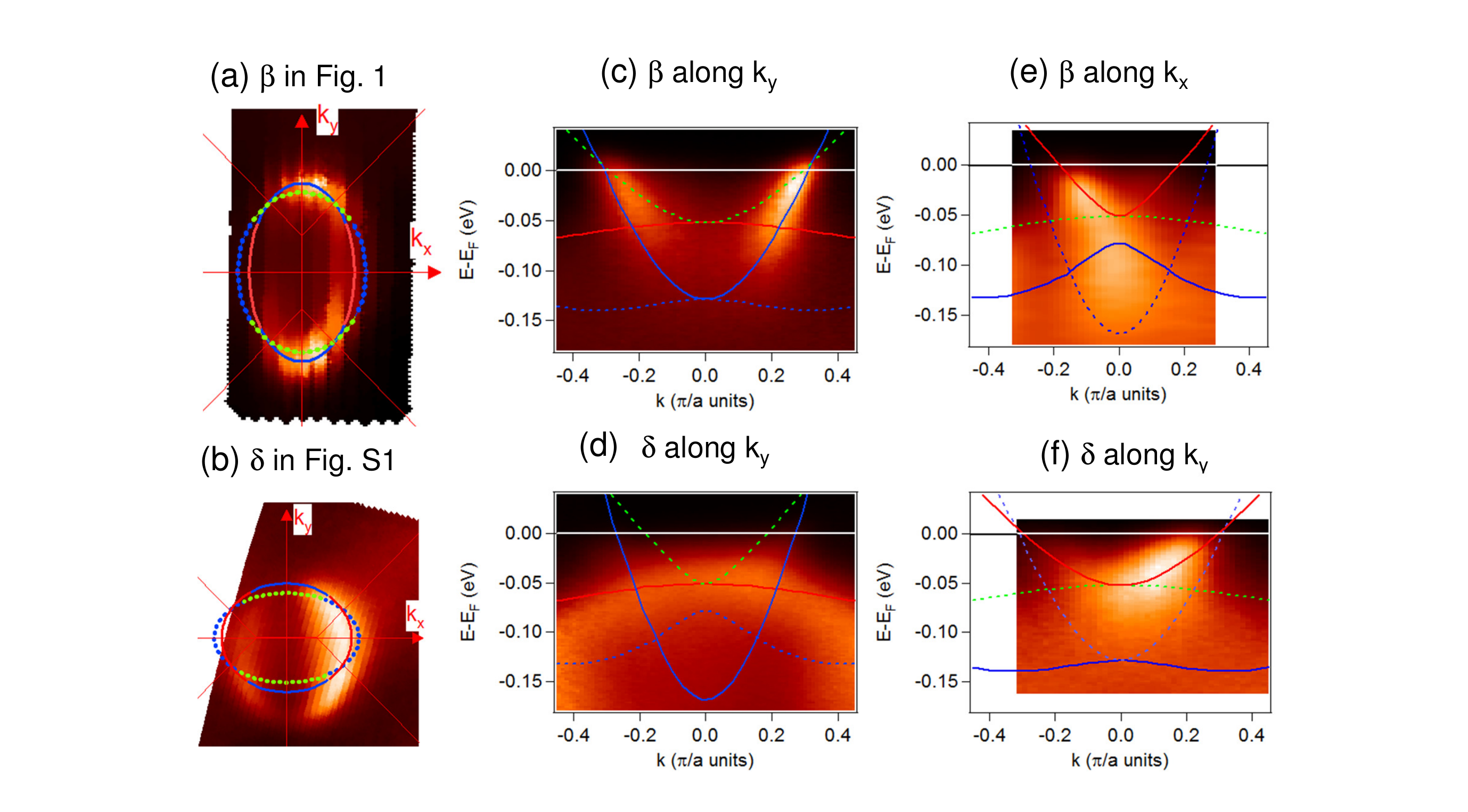}
\caption{(a-b) Zoom on the electron pockets even/$yz$ of panels $\beta$ and $\delta$ (Fig. 1 and \ref{FS_kz0}), together with sketches of the electron pockets. (c-f) Corresponding dispersions along $k_y$ (c-d) and $k_x$ (e-f), together with calculated bands rescaled by a factor 3. }
\label{Even}
\end{figure*}
\vspace{10mm}

\end{widetext}

\end{document}